\documentclass{article}
\usepackage[utf8]{inputenc}
\usepackage{authblk}
\usepackage{setspace}
\usepackage[margin=1.25in]{geometry}
\usepackage{graphicx}
\graphicspath{ {./figures/} }
\usepackage{subcaption}
\usepackage{amsmath}
\usepackage{lineno}

\usepackage[style=ieee, 
citestyle=numeric-comp,
sorting=none]{biblatex}
\title{Photonic Spiking Neural Networks with Highly Efficient Training Protocols for Ultrafast Neuromorphic Computing Systems}

\author{Dafydd Owen-Newns}
\author{Joshua Robertson}
\author{Mat\v{e}j Hejda}
\author{Antonio Hurtado}

\affil[1]{Institute of Photonic, SUPA Department of Physics, University of Strathclyde, Glasgow, UK.}
\affil[*]{Corresponding author. Email: dafydd.owen-newns@strath.ac.uk}

\date{}

\begin{document}

\maketitle

\begin{abstract}

Photonic technologies offer great prospects for novel ultrafast, energy-efficient and hardware-friendly neuromorphic (brain-like) computing platforms. Moreover, neuromorphic photonic approaches based upon ubiquitous, technology-mature and low-cost Vertical-Cavity Surface Emitting Lasers (VCSELs) (devices found in fibre-optic transmitters, mobile phones, automotive sensors, etc.) are of particular interest. Given VCSELs have shown the ability to realise neuronal optical spiking responses (at ultrafast GHz rates), their use for spike-based information processing systems has been proposed. In this work, Spiking Neural Network (SNN) operation, based on a hardware-friendly photonic system of just one Vertical Cavity Surface Emitting Laser (VCSEL), is reported alongside a novel binary weight 'significance' training scheme that fully capitalises on the discrete nature of the optical spikes used by the SNN to process input information. The VCSEL-based photonic SNN is tested with a highly complex, multivariate, classification task (MADELON) before performance is compared using a traditional least-squares training method and the alternative novel binary weighting scheme. Excellent classification accuracies of $>$\,94$\%$ are reached by both training methods, exceeding the benchmark performance of the dataset in a fraction of processing time. The newly reported training scheme also dramatically reduces training set size requirements as well as the number of trained nodes ($\leq$\,1$\%$ of the total network node count). This VCSEL-based photonic SNN, in combination with the reported 'significance' weighting scheme, therefore grants ultrafast spike-based optical processing with highly reduced training requirements and hardware complexity for potential application in future neuromorphic systems and artificial intelligence applications.

\end{abstract}





\section{Introduction}
\subsection{Photonic ANNs and Spiking VCSELs}
\label{Sect:1.1}
Artificial Neural Networks (NNs) are more frequently appearing in the fast-developing field of Artificial Intelligence (AI) thanks to their proven ability to provide high level performance at numerous complex information processing tasks \cite{Alzubaidi2021,Kazanskiy2022} such as computer vision and data classification. NNs, that draw inspiration from the networks of biological neurons in the brain, are formed of highly parallel structures of nodes (neurons) that realize numerous non-linear transformations to achieve efficient processing and decision making. The desire to move towards a more-than-Moore era of computing and diverge from traditional computing paradigms has also spurred on reports of brain-inspired computing, where now multiple systems of large-scale electronic ANN systems, such as the Loihi \cite{Davies2018}, TrueNorth \cite{DeBole2019} and BrainScaleS \cite{Pehle2022}, amongst others, exist. However, recently the number of works demonstrating optical approaches to beyond-Von Neumann computing is also growing, given the appeal of the inherent properties of the optical medium. Optical-based systems can operate with very low component crosstalk, perform across large data bandwidth ranges, and utilise non-interacting frequencies of lights to achieve highly parallel division-multiplexing applications \cite{Brackett1990}. More importantly however, the photonic platform can allow NNs to operate with low power consumption and ultrafast speed, overcoming the squeeze of power efficiency and clock speed faced by state of the art chip-scale electronic components \cite{Prucnal2017,Miller2017}. The realisation of light-based NNs, or so called Photonic Neural Network (PNN) systems, is therefore of key interest to all future AI applications where the benefits of increased operation speed and power efficiency is directly felt in the training and processing of large data volumes.

Despite the relative infancy of brain-inspired (neuromorphic) photonic systems, the field is advancing quickly with the attractive properties of optics already inspiring the production of multiple PNN accelerators \cite{Feldmann2019,Feldmann2021,Xu2019,Tait2017,DeLima2019,Mehrabian2018,Ashtiani2022,ZhangH2021}. These system, based on technologies such as phase change materials \cite{Feldmann2019,Feldmann2021}, optical modulators \cite{Xu2019,ZhangH2021}, and micro-ring weight banks \cite{Tait2017,DeLima2019,Mehrabian2018,Ashtiani2022}, amongst others, have risen to demonstrate different information processing tasks with photonic components. Neuromorphic photonic systems, built with semiconductor laser (SL) technologies, have also demonstrated intriguing neuron-like non-linear dynamics key to the operation of PNN systems \cite{Chen2022,Prucnal2016}. Vertical-Cavity Surface-Emitting lasers (VCSELs), well-established, commercially available and increasingly ubiquitous devices (found in mobile phones, data centres, supermarket barcode scanners, etc.), are one such neuromorphic SL technology. The ability of VCSELs to imitate the behaviours of biological neurons was first proposed in \cite{Hurtado2015}, where it was shown that controllable spike activation could be achieved with incoming optical signals by exploiting non-linear dynamics surrounding the injection locking condition. The neuron-like spiking responses were achieved at ultrafast (GHz) rates with spike widths of approximately 100\,ps, making the spiking responses $>6$ orders of magnitude faster than biological neurons, and multiple orders of magnitude faster than some electronic spiking systems. VCSELs have since showcased several common neuronal behaviours (such as integrate and fire spiking \cite{Robertson2019-Towards,Robertson2020-pattern-class}, refractoriness and rate coding \cite{Hejda2020}) in experimental realisations. Further, early demonstrations of neuromorphic information processing functionalities with VCSELs has been achieved using the spiking neuronal behaviours of VCSELs both experimentally, including image processing (edge feature detection) \cite{ROBERTSON2020-edge-det,Zhang2021Binary,Robertson2020Image}, pattern recognition \cite{Robertson2020-pattern-class} and exclusive OR (XOR) operation \cite{Zhang2021Pyramidal}, as well as theoretically through simulations of the Yamada and Spin-flip models (see \cite{Xiang2021-review} for a review). Moreover, single VCSEL devices have shown processing functionality when operated in combination with software-based Spiking Neural Networks (SNNs), achieving high classification accuracy at the MNIST hand written digit recognition task \cite{Robertson2022Ultrafast}. VCSELs therefore represent an increasingly promising technological approach to spike-based PNNs that operate with ultrafast, telecommunication wavelength, optical signals, that is hardware friendly, low power and low cost. 

Beyond the photonic imitation of spiking neuronal behaviours, the technique reservoir computing (RC) has been demonstrated as a powerful method for creating PNNs, yielding excellent performance in complex tasks while also benefiting from highly hardware-friendly architectures. First developed in the early 2000's \cite{Jaeger2001,Maass2002}, it was shown that in RC architectures (NN architectures with fixed-weight hidden layer connections) only the output layer required training to achieve successful performance. In these RC architectures, the nodes within the hidden layer are referred to as the "reservoir". The reservoir is therefore formed of unknown, fixed-strength, interconnected non-linear elements that are coupled to an output layer. Given the fixed nature of the connections, RC limits the training requirements of NNs, helping significantly reduce the resources (computational power and time) needed to train large networks of nodes for successful operation \cite{Jaeger2001}. There have been numerous theoretical \cite{Hulser2022,Rohm2020} and experimental \cite{Vandoorne2014,Brunner2013,Vinckier2015} reports of reservoir computers based on the photonics platform with VCSELs again featuring frequently \cite{Bueno2017,Argyris2019,Vatin2018,Vatin2019,Vatin2020,Bueno2021}. VCSELs are one of the photonics devices where both time-delay reservoirs (TDRs) \cite{Vatin2018,Vatin2019,Vatin2020,Bueno2021}, which multiplex nodes over time to construct virtual networks and create memory through feedback connections, and spatial temporal reservoirs \cite{Porte2021,Skalli2021}, which spatially multiplex nodes, have demonstrated successful operation. These systems have revealed impressive performance on numerous complex processing tasks, utilising both off-the-shelf \cite{Bueno2021}, and bespoke large area (LA)-VCSEL designs \cite{Porte2021}, to realise photonic computing with continuous (non-spiking) VCSEL signals (see \cite{Skalli2022} for a review). More recently, we demonstrated for the first time the combination of the reservoir computing technique with a neuromorphic spiking VCSEL-neuron \cite{Owen-Newns2023}. In that report we revealed that by using the time-multiplexing technique and the masking of inputs, we could interpret a single VCSEL neuron as an fixed-weight interconnected virtual SNN. This photonic system realised both the reduced training requirements of an RC system as well as the sparse, all-or-nothing, binary spike-based representation of a full SNN. The VCSEL-based SNN operated with short temporal nodes (250\,ps), fast optical spikes (100\,ps-long), and a configurable network node number, that allowed it to demonstrate successful operation on the benchmark Iris Flower classification task \cite{Fisher:1936} with excellent performance. 

In this work, we show that we can not only use time-multiplexing to create a VCSEL-based photonic SNN capable of a highly complex processing task, but more importantly we can, through alternative training schemes, achieve high performance with significantly reduced training requirements (namely reduced training set sizes and number of trained nodes). This paper adopts the following structure. Firstly, in Section\,\ref{Sect:1.2} we will introduce the experimental setup used in this work to create the photonic SNN and discuss its operating principles. Next, in Section\,\ref{Sect:1.3} we will introduce the experimental classification task and explain how data is prepared for injection into the VCSEL-based SNN. Then, in Section\,\ref{Sect:2.1} we will discuss the results of the classification task and explore training through the standard linear least-square regression method. In Section\,\ref{Sect:2.2} we will then introduce the alternative binary weight training scheme and discuss its performance relative for various training set sizes and node requirements. Finally, we will provide conclusions in Section\,\ref{Sect:3}.

\begin{figure}[!t]
\centering
\includegraphics[width=5.5in]{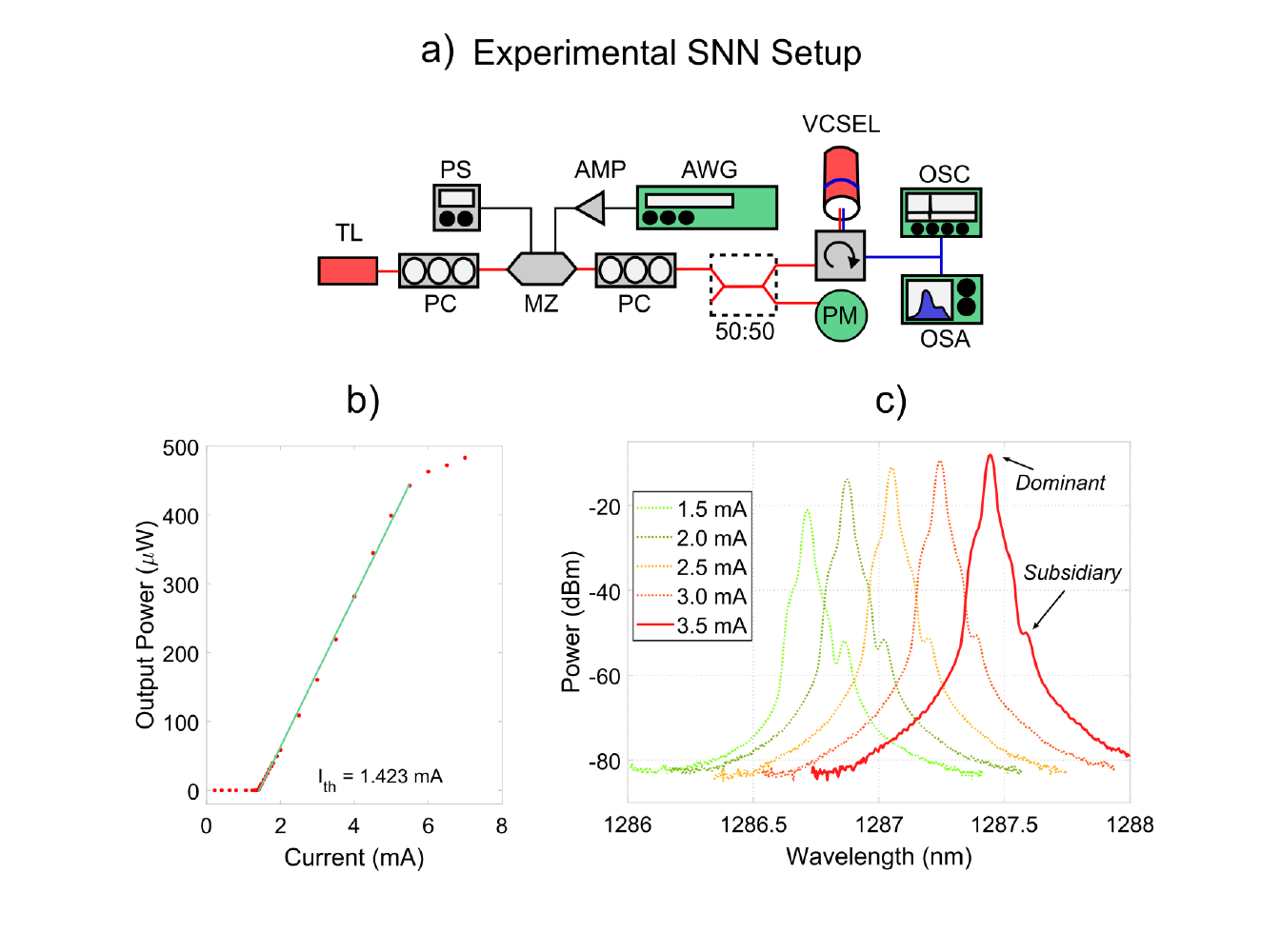}
\caption{a) The experimental setup used to create the photonic SNN with a single VCSEL device. b) The lasing threshold (Light-Current) characteristics of the experimental VCSEL-neuron reveals a threshold current of 1.423 mA. c) The spectra of the experimental VCSEL-neuron at different bias currents. At the operating current (3.5\,mA - red) the device exhibited two orthogonal polarisation peaks at 1287.44\,nm and 1287.59\,nm. Measurements were taken with a room temperature (293\,K) stabilised device.}
\label{Fig: Intro1}
\end{figure}

\begin{figure}[!t]
\centering
\includegraphics[width=6in]{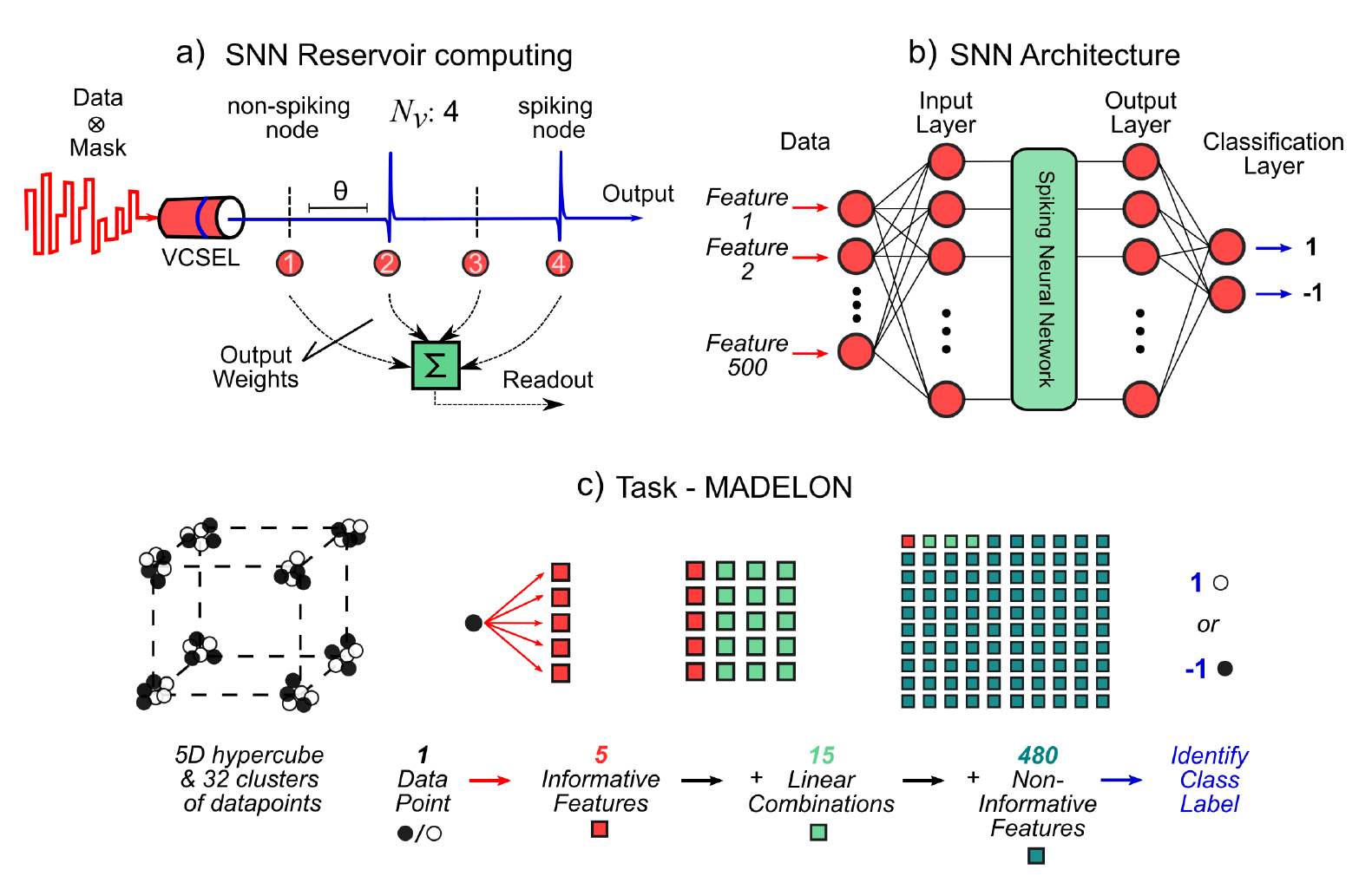}
\caption{a) The photonic SNN operating principle. Masked input data is injected into the VCSEL forming the photonic SNN. The input information is time-multiplexed with each $\theta$ time-slot representing a virtual node (neuron) in the SNN. The time-multiplexed output of the VCSEL is interpreted as a binary node output, either spiking (1) or non-spiking (0). The output layer nodes are weighted offline to provide the photonic SNN readout. b) The architecture of the photonic SNN. 500 features per datapoint are input into the SNN system. In this experimental demonstration the hidden layer was selected to be made up of either 2048 or 4096 virtual nodes. The output layer is trained offline to provide a datapoint classification (either +1 or -1). c) The MADELON classification task. Datapoints are placed on the corners of a 5-dimensional hypercube and randomly labelled -1 or +1. Each artificial data point has 5 informative features (each corresponding to 1 of 5 dimensions). 15 linear combinations of the original features are made creating 20 total features. 480 non-informative features (that carry no information of the label) are then added to the feature set. The task requires all 500 data features to be used to classify the initial data point label -1 or +1.}
\label{Fig: Intro2}
\end{figure}

\subsection{Photonic SNN Setup and Operation}
\label{Sect:1.2}
In this work, we report a hardware-friendly photonic SNN based on a single time-multiplexed VCSEL-neuron. The experimental setup used to build and investigate the photonic SNN architecture of this work is shown graphically in Fig.\,\ref{Fig: Intro1}\,(a). The photonic SNN is built with a single telecom-wavelength VCSEL and commercially-sourced fibre-optic telecom components. The selected VCSEL had wavelength emission within the key optical telecommunication O-band window (centred around 1300\,nm) to showcase full compatibility of the system with optical networking and data centre technology. In the experimental setup, external light injection is provided by a tuneable laser (TL) source which is intensity modulated by a 10\,GHz Mach Zehnder modulator (MZ). The MZ is responsible for encoding the input information in the intensity of the optical injection before it enters the VCSEL-neuron. The MZ is driven by a DC power supply (PS) and a fast 12\,GSa/s 5\,GHz arbitrary waveform generator (AWG) that generates the required input information. A 10\,dB electrical amplifier (AMP) is used to increase the signal from the AWG, and a polarisation controller (PC) is used to maximise TL light coupling into the MZ. The modulated injection is split by a fibre coupler to make a measurement of injection power before the TL light enters the circulator and the VCSEL. The output of the VCSEL is captured by the circulator and fed to analysis, where it was measured by an optical spectrum analyser (OSA) and a fast 40\,GSa/s 16\,GHz real-time oscilloscope (OSC). The threshold and spectral characteristics of the O-band VCSEL used in the photonic SNN system are shown in Figs\,\ref{Fig: Intro1}\,(b)\,\&\,(c). When stabilised at room temperature (293\,K), the VCSEL exhibited a threshold current of 1.423\,mA and produced two orthogonal polarisation peaks in the fundamental transverse mode. In this work, with the selected operating current of 3.5\,mA, these orthogonal polarisation peaks occurred at 1287.44\,nm and 1287.59\,nm, producing a total output power of $\sim$220 $\mu$W. Using a PC, optical injection was polarisation-matched to the subsidiary (1287.59\,nm) polarisation mode of the VCSEL, with injection made using a negative frequency detuning to induce polarisation switching and injection locking. The modulating intensity is responsible for the triggering of high speed (approx. 100\,ps-long) neuron-like spiking responses near the injection locking boundary.

In this work, as in \cite{Owen-Newns2023}, we combine the concepts of reservoir computing with the photonic spiking VCSEL-neuron to create a fully photonic SNN using significantly reduced hardware (a single laser device). As discussed in Section\,\ref{Sect:1.1}, reservoir computers are a type of artificial NN that host a hidden layer of interconnected nodes whose connections and weights are fixed. This means that in reservoir computers only the output layer of the network needs to be trained to achieve high performance \cite{Jaeger2001}. Further, reservoir computers can make use of time-multiplexing to interpret the output of one node as the output of many nodes, by sampling at set times ($\theta$). Here, using the VCSEL-neuron, we apply the same concepts to interpret the output of one spiking VCSEL as an entire photonic SNN, see Fig\,\ref{Fig: Intro2}\,(a). The output of the VCSEL is time-multiplexed at intervals of $\theta$ to produce a controllable number of network nodes ($N_v$). The coupling of network nodes is achieved given the duration of the nodes ($\theta$) is less than the timescale of the neuron-like spiking non-linearity in the VCSEL (typically $\sim$\,1\,ns \cite{Hejda2020}). This condition allows the neuronal leaky integrate-and-fire and refractory behaviours of the VCSEL-neuron to link multiple nodes coupled in time, forming a complex interconnected virtual network structure. As in \cite{Owen-Newns2023} (and those of other VCSEL-based reservoir computers \cite{Bueno2021}), the input data is prepared by randomly masking data point features and injecting them into the VCSEL as continuously modulated light. Following injection, each $\theta$-long node will deliver an all-or-nothing binary output, determined by the presence (or absence) of a fast 150\,ps optical spike, effectively realising the output of a fast photonic SNN. In this system the output weights (the weights applied to each node) are trained and calculated offline to provide the readout of the system for the prescribed processing task (the task and training schemes are discussed in Sections\,\ref{Sect:1.3}\ \& \ref{Sect:2}). 

In \cite{Owen-Newns2023}, we recently demonstrated experimentally that a VCSEL-neuron system can successfully realise SNN operation with fast optical spiking signals. There we showcased the benchmark nonlinear Iris flower classification task, comprising of 150 non-linearly separable data points (flowers), 4 features per data point (sepal \& petal length/height) and 3 flower classes. Yet, using our VCSEL-based photonic SNN we demonstrated that a very high overall average classification accuracy of $>$97\,\% could be achieved, showcasing the photonic SNNs powerful flexibility and performance. Importantly, in that work the training of the output weights was achieved using an ordinary least squares regression method, and hence the node outputs were treated as if it were continuous (spiking nodes were assigned the value 1.0, and others given 0.0). This result required the offline calculation of all node weights as real number (floating point) values. In this report we focus now on completing a much more complex problem and apply an alternative training scheme, whereby the discrete nature of the binary all-or-nothing spiking responses can be used to improve the speed and efficiency of training. In the following sections we will introduce the new, high-complexity, non-linear classification task (MADELON) \cite{GUYON20071438} and demonstrate successful performance of the photonic SNN with two training methods; the aforementioned least-square regression training method, and an alternative node-significance training method, to highlight the attainment of notable improvements in both training speed and resource requirements. 

\subsection{Data Set Preparation}
\label{Sect:1.3}
We further demonstrate the capability of the VCSEL-based photonic SNN by applying the system to a highly complex classification task. The task used in this work is an artificial dataset named 'MADELON', which was created to test feature selection methods as part of a NIPS 2003 challenge \cite{GUYON20071438}. A simplified schematic of the MADELON data creation process is shown in Fig.\,\ref{Fig: Intro2}\,(c). The dataset is made up of 32 clusters of datapoints placed on the vertices of a 5-dimensional (5D) hypercube. Two classes of data are created by randomly labelling each datapoint with the value -1 or +1. Each data point has 5 informative features (one for each dimension) that can be used to identify the class of the data. Using the 5 informative features, 15 linear combinations are created, increasing the feature set size to 20. Finally, the 20 features are randomly scattered between 480 non-informative (no predictive power) features creating a total of 500 features per data point. The task is therefore a multivariate, nonlinear, two-class problem with continuous input variables. The task is of high complexity given a classification must be made using a feature set where no feature by itself is informative. The problem is further complicated by the addition of non-informative features, which require the system to eliminate and ignore distractions. 

The entire MADELON dataset contains 2000 data points intended for training. However, here to demonstrate the operation of the VCSEL-based photonic SNN, we used 300 data points in total (150 of each class). Limiting the number of input points reduced the length of the time-multiplexed output generated during the experimental runs, hence allowing us to test different network architectures with diverse node numbers while remaining within the memory of our experimental equipment. As the MADELON task is of much higher complexity than the preliminary Iris Flower classification task \cite{Owen-Newns2023}, a higher number of virtual nodes ($N_v$) were selected, namely 2048 and 4096, to test the classification performance of the photonic SNN. 

The photonic SNN architecture used to complete the task is shown in Fig.\,\ref{Fig: Intro2}\,(b). Each of the 500 features are fed into the $N_v$ input layer nodes via a random mask. The masking of the input data is achieved by multiplying a column vector of all 500 features with a 500\,x\,$N_v$ random matrix, resulting in a vector of $N_v$ components. The masked data is fed into the hidden layer of the photonic SNN, where fixed connections and weights are created by the non-linear temporal dynamical behaviour in the VCSEL. Experimentally, the randomly masked data is generated by the AWG and modulated by the MZ into the optical injection of the VCSEL. The output layer of $N_v$ nodes is then provided by the segmentation of the VCSEL's spiking output timeseries at $\theta$ intervals. In this work a node separation ($\theta$) of 250\,ps was used to time-multiplex network nodes, creating multiple interacting nodes within the 1\,ns timescale of non-linearity in the VCSEL \cite{Hejda2020,Robertson2020-pattern-class}. The final classification layer of the SNN then provides a -1 or 1 label to the data following the offline training and weighting of output nodes. The following section will discuss the two training methods implemented in this work.


\section{Results}
\label{Sect:2}

First, the MADELON task was run in the photonic VCSEL-based SNN using a network node count of $N_v=2048$. This results in a total processing time of 512\,ns per datapoint, in the photonic SNN. The VCSEL was driven with a current of 3.5\,mA ($\sim$\,2.5 times the lasing threshold current), at a temperature of 293\,K; the injected light had mean power of 142.6\,$\mu$W, and was injected into the VCSEL with an initial frequency detuning of -5.4\,GHz with respect to the resonant wavelength of the VCSEL's subsidiary polarisation mode peak. These conditions put the VCSEL into a state in which it was injection-locked to the externally-injected signal. Here, near the injection locking boundary, optical spikes could be triggered by the varying intensity of the modulated optical injection light.

On a second experimental run, the number of virtual (spiking) nodes in the network was increased to $N_v=4096$ (1024\,ns processing time per datapoint). In this second run the VCSEL was biased with a current equal to 3.45\,mA, at a set temperature of 293\,K. The injected light signal, encoded with the input data, had a mean optical power of 161\,$\mu$W and was injected with an initial frequency detuning of -4.6\,GHz (with respect to the resonance of the VCSEL's subsidiary polarisation mode).

The output of the photonic SNN, for a given datapoint, is a vector $S$ of length $N_v+8$. The 8 additional virtual nodes are the result of zero padding at the end of the input data sequence, equivalent to a 2\,ns reset of the system between inputs. Resetting the system between consecutive datapoint inputs prevented the previous datapoint from influencing the new datapoint and generating undesired spike activations. In the SNN output vector $S$, the i\textsuperscript{th} element was readout as 1 if the i\textsuperscript{th} node spiked within the 250\,ps-long time-multiplexed segment, and was assigned value 0 otherwise.

\begin{figure}[!t]
    \centering
    \includegraphics[width=4in]{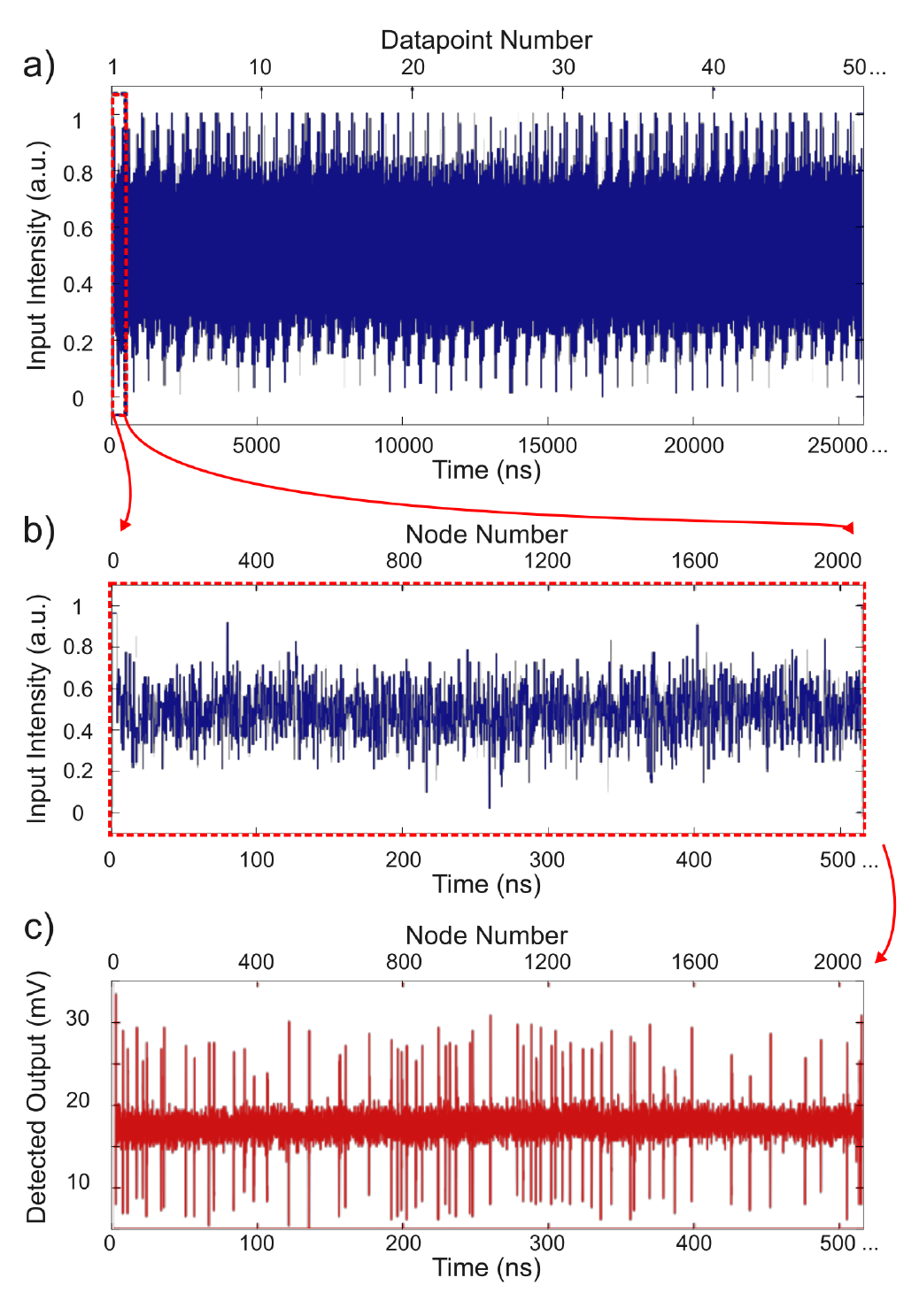}
    \caption{Time series showing the measured input (blue) and output (red) of the photonic SNN for a network architecture of 2048 virtual nodes. a) A section of injection waveform showing the time-multiplexed intensity variations of 50 consecutive masked datapoints of the MADELON dataset. b) A 512\,ns segment of the input timeseries corresponding to the first MADELON datapoint. Each of the 2048 virtual nodes have a duration of 250\,ps, creating the 512\,ns input timeseries. c) The corresponding measurement of the SNN output when injected with the first MADELON datapoint.}
    \label{fig:time_series}
\end{figure}

Figure \ref{fig:time_series} shows experimentally-measured time-series depicting the optical input (Fig.\,\ref{fig:time_series}\,(a-b)) and output signals (Fig.\,\ref{fig:time_series}\,(c)) of the photonic SNN. The example shown in Fig.\,\ref{fig:time_series} is for a network architecture with 2048 nodes. Time-multiplexing is used to inject each masked datapoint into all the (virtual) nodes of the photonic SNN (at a rate of 250\,ps per virtual node). This creates a time varying signal for each datapoint, which is applied as an intensity modulation of the light injected into the SNN. During the experiment, all 300 masked datapoints (time varying signals) are injected into the SNN sequentially, creating a large optical input timeseries. Figure \ref{fig:time_series}\,(a) depicts part of the large optical input time-series, showing 50 sequentially encoded masked datapoints of the MADELON dataset. Figure \ref{fig:time_series}\,(b) plots the first 512\,ns of the optical input time-series in Fig.\,\ref{fig:time_series}\,(a), which corresponds to the input signal encoding the first MADELON datapoint. As the amplitude of the optical input signal varies, it will trigger the firing of optical spike events from the VCSEL at specific times (virtual nodes). Therefore, different input signals (masked datapoints) entering the photonic SNN will elicit different optical spike trains from the network. The optical spiking time-series, measured at the output of the SNN in response to the first injected MADELON dataset point (input signal in Fig. 3b), is shown in Fig.\,\ref{fig:time_series}\,(c).

\begin{figure}[!t]
    \centering
    \includegraphics[width=5in]{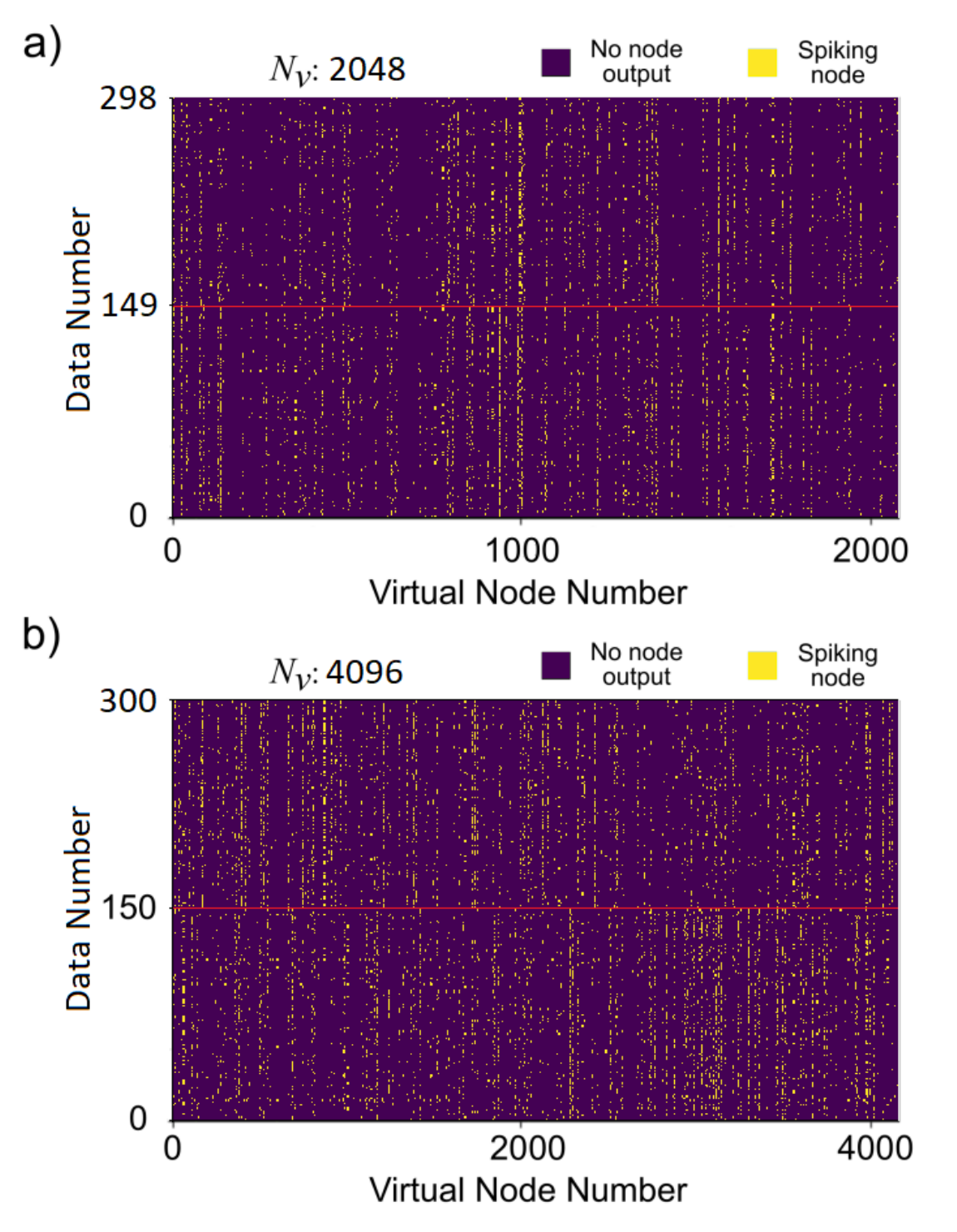}
    \caption{Temporal maps showing the optical spiking patterns produced by the SNN in response to the input dataset using a) 2048 virtual nodes and b) 4096 virtual nodes.}
    \label{fig:spike_mesh}
\end{figure}

Figure \ref{fig:spike_mesh} plots a 2D temporal map that collects in one plot the spiking output from all virtual nodes of the photonic SNN for all 300 consecutively injected data points in the MADELON task. In Fig.\,\ref{fig:spike_mesh} a yellow (blue) dot indicates a spiking (non-spiking) response from the VCSEL-based SNN, and the resulting 1 (0) binary output of the virtual node in vector $S$. The output vector $S$ is plotted horizontally for each input datapoint, revealing trends in node activation across inputs and classes. The results for photonic SNNs configured with 2048, and 4096, virtual (spiking) nodes are plotted in Figures \ref{fig:spike_mesh}\,(a) and (b), respectively. The red line through the centre of the map separates the results for each distinct class (-1 \& +1) of input datapoint. Figures \ref{fig:spike_mesh}\,(a) and (b) show that each input datapoint of the same class elicits a similar spike train at the output of the photonic SNN. These spike trains are sufficiently distinct from those obtained by the other class of datapoint, to allow for a classification operation.

For the case of the SNN with a 2048 node architecture (Fig.\,\ref{fig:spike_mesh}\,(a)), a total of 298 datapoints (149 per class) were used in the experimental run. The reason for this is that the first datapoint in class $-1$ was affected by an experimental artifact, and was hence remove from analysis. One point from class $+1$ was therefore also removed to keep the size of each set equal, resulting in a total of 298 points. For the network architecture with $Nv=4096$ (Fig.\,\ref{fig:spike_mesh}\,(b)), all 300 data points (150 per class) of the MADELON task were used to test the network operation.

Training is completed using a randomly selected subset of the input datapoints, and their corresponding spiking responses. The randomly selected subset contains an equal number of points ($N_t$) of each class, for a total training set size of $2N_t$. In our photonic SNN, the training is used to calculate the weights applied to the output layer nodes (i.e. the temporal spiking patterns produced by the network and the corresponding vector $S$). The output weights are determined in such a way to achieve the best performance on this set, with the assumption that the training set is a good representation of the entire task dataset. The input datapoints unused in the training process are subsequently used to test the performance of the MADELON task. The output layer weights calculated during training are applied to the output layer to infer the class of a given test datapoint. The inferred datapoint class is then compared to the true class label and performance is measured using the classification accuracy, the fraction of correctly classified datapoints across all tested datapoints. The training/testing process was repeated several times, randomly selecting the training set each time in order to find an average value for the performance (a process known as random cross-validation).

\subsection{Ordinary Least Squares}
\label{Sect:2.1}

In this work, we first used the Ordinary-Least-Squares (OLS) training method to test the performance of the photonic SNN in the MADELON task, a standard training method applied in other photonic RC systems \cite{Bueno2020,Owen-Newns2023}. Here, the output spiking patterns from the VCSEL (and corresponding vector $S$) are treated as a binary sequence of floating-point 1.0s and 0.0s. Then, the output layer weights are found via linear regression where the values of each node (1.0 or 0.0) are arranged into matrix $X$ (where the rows of $X$ are the output vectors $S$). The labels (expected values) of each datapoint form a second column vector $Y$. The weight calculation then consists of finding the weight matrix $W$ that minimises Equation\,(\ref{eq:error}):

\begin{equation}
    E_{rror}(W)=\left|Y-XW\right|^2.
    \label{eq:error}
\end{equation}

Equation\,(\ref{eq:error}) is solved using $W=\left(X^TX\right)^{-1}X^TY$ (the Penrose inverse of $X$ multiplied by $Y$). For this task there are two output classes (-1 \& +1), so the labels are two-element vectors, in which the n\textsuperscript{th} element is set to $1$ to denote the n\textsuperscript{th} class, with the other elements being zero. This also means the resulting weight matrix $W$ has dimension of $\left(N_v\times2\right)$. Using $W$, the prediction for the class of a datapoint (produces a spike pattern $S$) can be found by taking $argmax(S*W)$. Applying the weights to the spike pattern vector gives a score towards each class label, where the class with the highest score is the inferred label (prediction) of the photonic SNN system.

\begin{figure}[!t]
\centering
\includegraphics[width=5.5in]{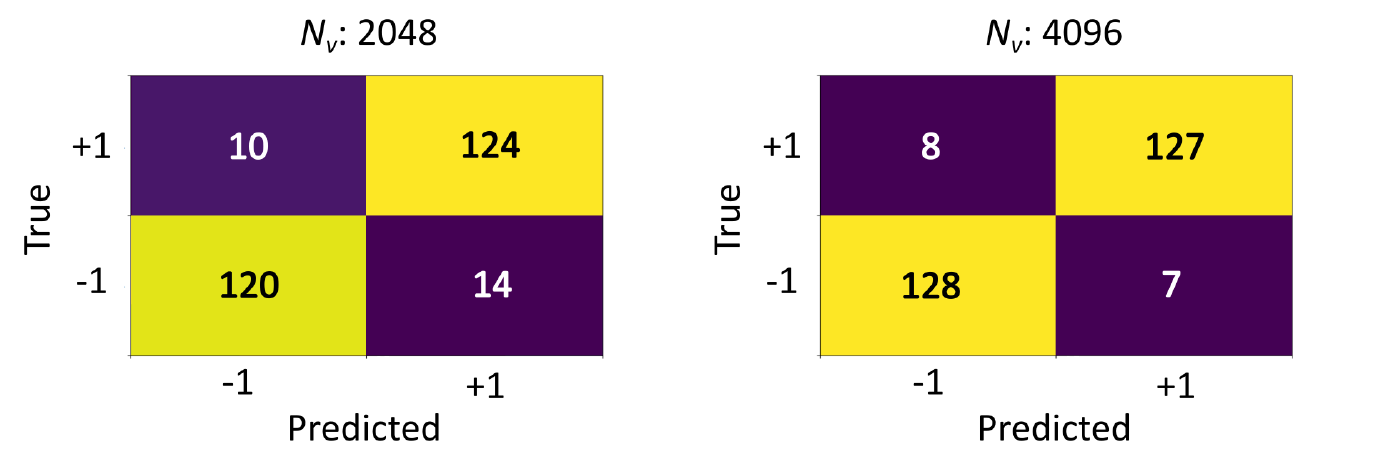}
\caption{Confusion matrices depicting the performance of the photonic SNN when trained using the ordinary least squares method. A training set size of 15 ($N_t=15$) was used to achieved accuracies of 91\,$\%$ and 94.4\,$\%$ for architectures of 2048 and 4096 nodes, respectively.}
\label{fig:ols-conf}
\end{figure}

Using only 15 training datapoints per class (10\,$\%$ of the total 300 input datapoints), the OLS training method achieved very high performance with a peak accuracy of $91\%$ (for a 2048 node architecture) and $94.5\%$ (for a 4096 node architecture). Example confusion matrices for $N_v=2048$ and $N_v=4096$, trained using this method and $N_t=15$ are shown in Fig.\,\ref{fig:ols-conf}. As expected, the network architecture with the higher node number ($N_v$) provide the better classification performance overall. The larger number of nodes creates more spiking responses from the VCSEL, allowing for more class-specific responses and a better classification. The capability to adjust node number at will is an inherent advantage of the photonic SNN of this work, as no physical/hardware changes are required to increase the network architecture. Instead, all that is require is the alteration of the input mask and a longer timeseries measurement. This makes the reported VCSEL-based photonic SNN an attractive and flexible system for the fast, optical spike-based, implementation of numerous processing tasks.  

\subsection{Binary Weight Training}
\label{Sect:2.2}

As discussed in the previous section (Section \ref{Sect:2.1}) the standard OLS training method can deliver high classification performance during the MADELON task. However, the OLS method does not take full advantage of the discrete binary nature of the fast optical spiking responses in the photonic SNN. For example, it can be observed that in the ideal case of an SNN response, spiking patterns would be perfectly consistent between data of the same class, as well as completely distinct from data of alternative classes. In this ideal case, $2^k$ classes would only require the use of $k$ nodes to be complete separate and identify datapoints. In the ideal case of the MADELON task this would mean only 1 node per class is required for identification. In the case of the experimental data, initally identifying nodes that have more predictive power (i.e. spike only for a particular class) prior to training, would therefore allow for creation of weight matrices that retain high classification accuracy, despite the use and calculation of only a few nodes (a fraction of the total node count).

We introduce the following process for finding weights for this SNN:
\begin{enumerate}
    \item Count the number of spikes that occurred for each class in each node, denoted $s_{n,i}$ ($\#$ spikes from data of class $i$ occurring in node $n$).
    \item Calculate the significance score of each node towards each class according to Equation\,(\ref{eq:Significance}):
    
    \begin{equation}
    z_{n,i}=s_{n,i}^2/\sum_is_{n,i}
    \label{eq:Significance}
    \end{equation}
    
    \item For each class, choose the top $N_n$ scoring nodes and set their weight towards that class ($W_{n,i}$) to 1, and all remaining node weights to zero.
\end{enumerate}

\begin{figure}[!t]
    \centering
    \includegraphics[width=6in]{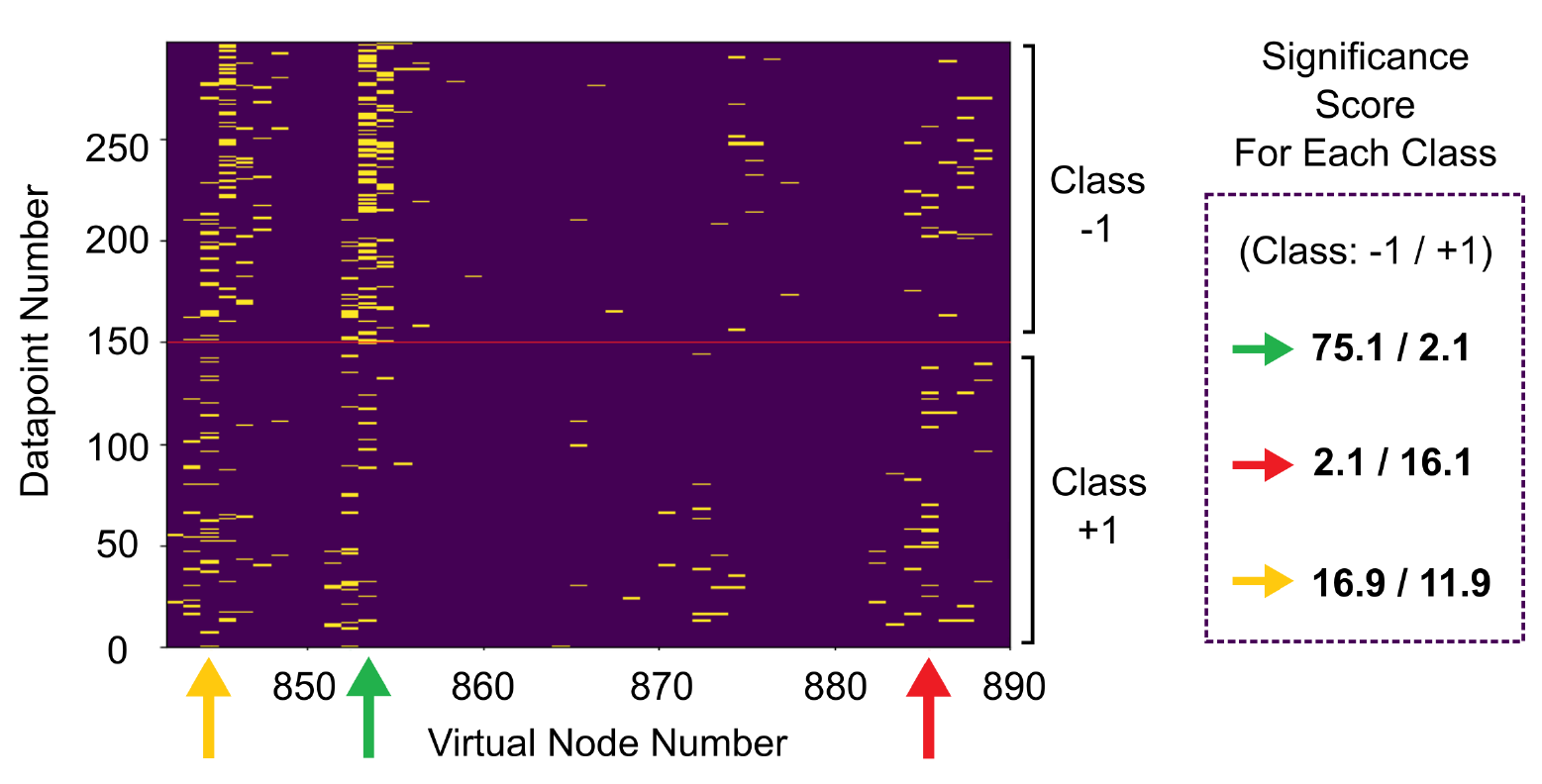}
    \caption{Section of a 2D temporal map for a network architecture of 2048 nodes (between nodes 842 and 890). The arrows mark nodes 844 (orange), 853 (green) and 885 (red). In node 853, 88 spikes occur for class -1, and 15 for class +1, giving it a high significance score of 75.1 (for class -1) and 2.1 (for class +1). Node 844 has significance scores of 16.9 (for class -1) and 11.9 (for class +1). Node 885 has significance scores of 2.1 (for class -1) and 16.1 (for class \+1). Node 853 has the highest significance score meaning it would likely be used to perform a classification.}
    \label{fig:spikes-zoom-sig}
\end{figure}

Applying the aforementioned algorithm provides the 'significance scores' for all network nodes, with higher scores obtained for nodes that contain not only more spikes, but nodes that have their spikes more concentrated in one specific class. The resulting output weight matrix $W$ will only have nonzero entries on nodes that are strongly indicative of one specific class. The 2D temporal maps shown in Fig.\,\ref{fig:spikes-zoom-sig} reveal the spike trains of a subset of network nodes, for the case when the photonic SNN was operated with a total node count of 2048. This plot demonstrates the significance scoring system applied to different selected nodes. For example, Fig.\,\ref{fig:spikes-zoom-sig} shows that node 844 (marked with an amber arrow) elicited an optical spike for 57 of the total 298 input datapoints used in this experimental run. In this case, the spiking responses from $n=844$ are evenly distributed across each class (31 spikes for class -1, and 26 for class +1). For this specific case, the significance scores for this node are $z_{844,-1}=16.9$ (class -1) and $z_{844,+1}=11.9$ (class +1). In parallel, Fig.\,\ref{fig:spikes-zoom-sig} shows that Node 885 (marked with a red arrow) fires fewer optical spikes overall. In total 30 responses are elicited, however because they are less evenly distributed (and hence more indicative of a particular class), the node scores $z_{885,-1}=2.1$ (class -1) and $z_{885,+1}=16.1$ (class +1), a class +1 significance score similar to node 844. Finally, when the appearance of fast optical spikes is much more biased towards one specific class, and occur more likely overall, the resulting significance score $z$ is much higher. For example, in node 853 (marked with a green arrow) a total of 103 spikes occurred, with 88 of those occurring for class -1, yielding scores of $z_{853,-1}=75.1$ and $z_{853,+1}=2.1$. With this method, the weight $W_{853,-1}$ of node 853 would be set to 1, as if a spike is detected in node 853, it is likely that the datapoint is of class -1. In this example the other highlighted nodes would have their weights ($W_{844,-1}$ \& $W_{885,-1}$) set to zero.

For a data point to be classified the spike output vector $S$ must be multiplied by the weight matrix $W_{i}$ of each class. The class with the highest sum value is the prediction for the class of the data point. Since the output layer weights are binary, this is equivalent to applying a filter to the optical spike trains and counting the number of spikes remaining. Using this method, the performance can be tuned by choosing the number of nodes ($N_n$) that are selected to be used for training (training node number). Using a fraction of the total number of available nodes can result in equal (if not better) performance, while keeping the weight matrix and the number weight calculations, sparse.

\begin{figure}
    \centering
    \includegraphics[width=5.5in]{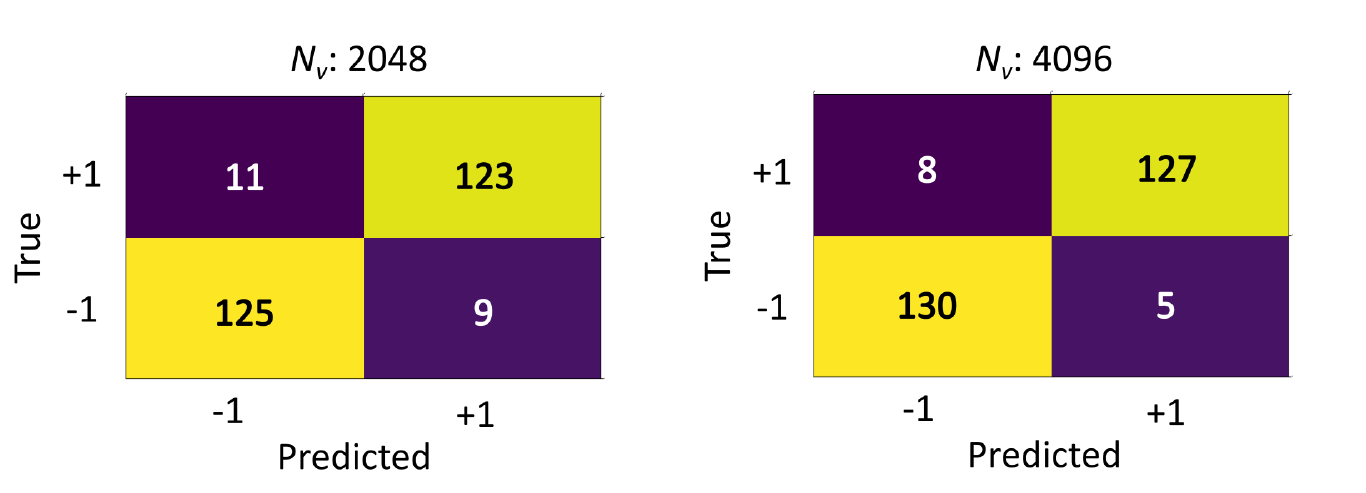}
    \caption{Confusion matrices depicting the performance of the photonic SNN when trained using the significance score method. A training set size of 15 ($N_t=15$) was used to achieved accuracies of 92.5\,$\%$ and 95.2\,$\%$ for architectures of 2048 and 4096 nodes, respectively. Only the 20 most significant nodes ($N_n=20$) were used to train the SNN.}
    \label{fig:sig-conf}
\end{figure}

\begin{figure}[!t]
    \centering
    \includegraphics[width=6in]{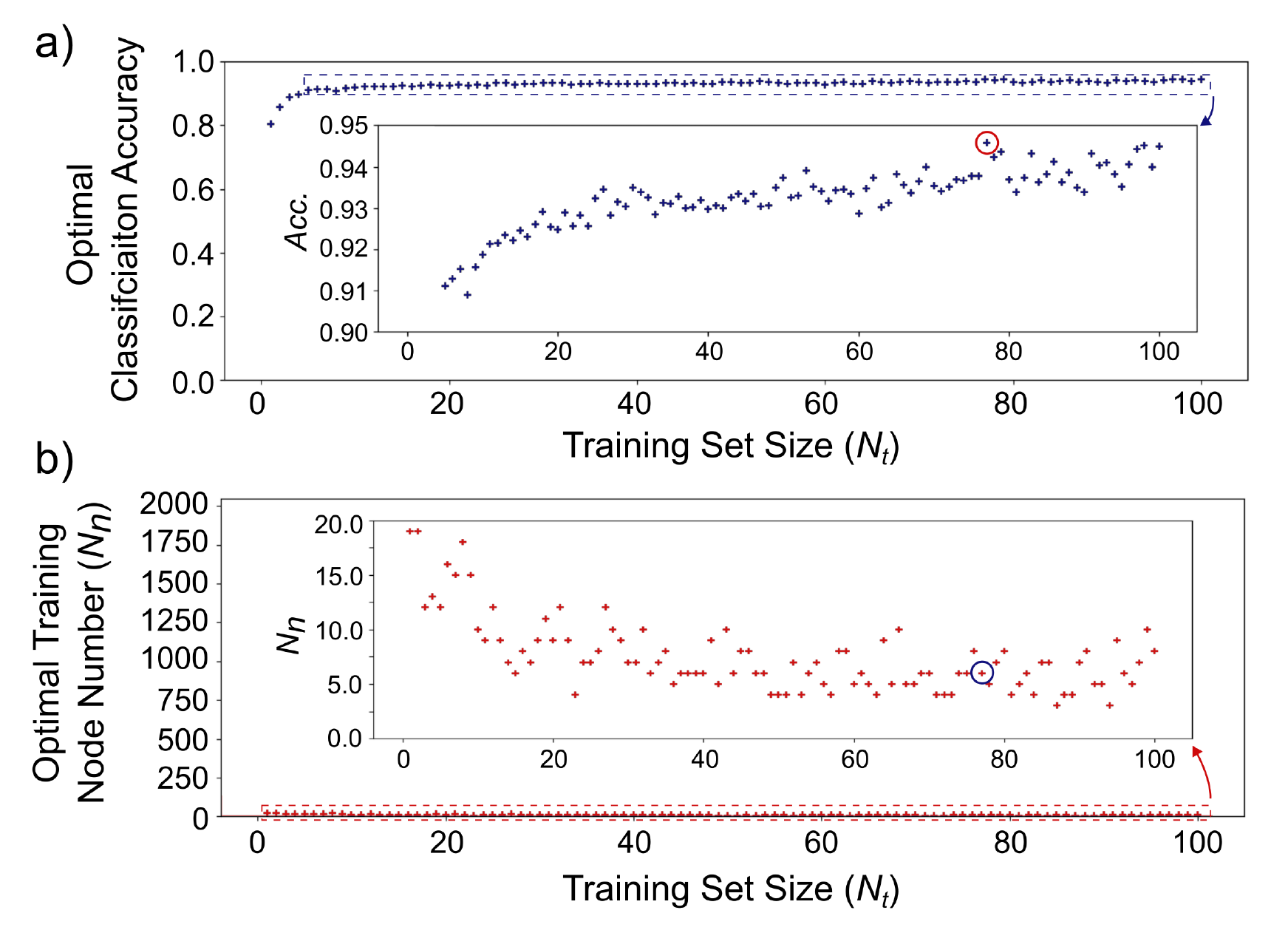}
    \caption{Photonic SNN performance versus the training set size, using a 2048 node architecture. a) The optimal classification accuracy (averaged over 10 random selections of the training set) is plotted against training set size ($N_t$, blue). The inset highlights the consistently high performance over various training set sizes ($N_t$ ranges from 1 to 100). b) The number of training nodes ($N_n$) used to achieve the corresponding optimal accuracy, plotted against increasing training set size (red) . The peak accuracy, 94.4$\%$, occurs when $N_n=6$ and $N_t=76$ (circled).}
    \label{fig:perf_test_2}
\end{figure}

\begin{figure}[!t]
    \centering
    \includegraphics[width=6in]{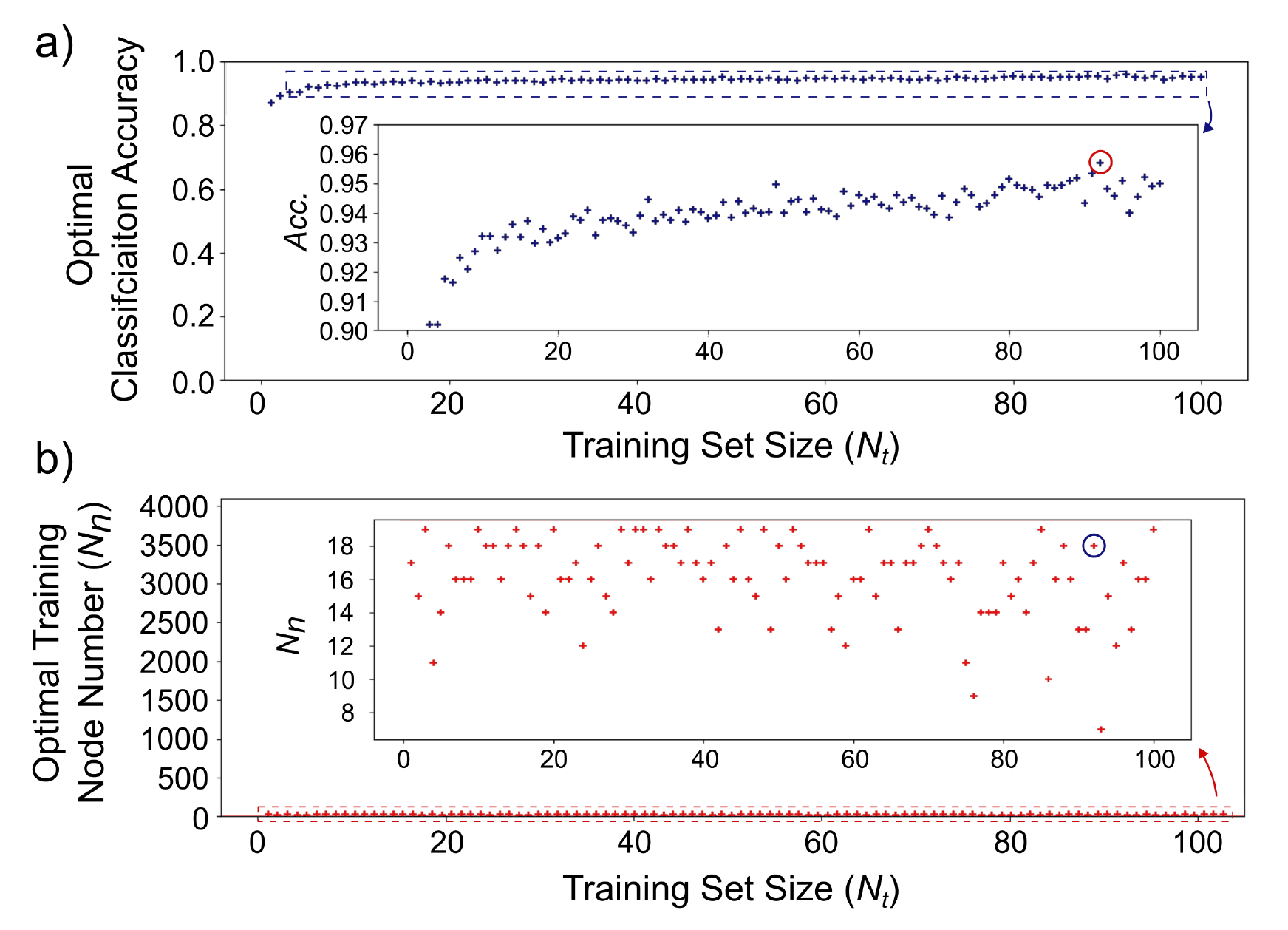}
    \caption{Photonic SNN performance versus the training set size, using a 4096 node architecture. a) The optimal classification accuracy (averaged over 10 random selections of the training set) is plotted against training set size ($N_t$, blue). The inset highlights the consistently high performance over various training set sizes ($N_t$ ranges from 1 to 100). b) The number of training nodes ($N_n$) used to achieve the corresponding optimal accuracy, plotted against increasing training set size (red). The peak accuracy, 95.7$\%$, occurs when $N_n=18$ and $N_t=91$ (circled).}
    \label{fig:perf_test_4}
\end{figure}

Figure\,\ref{fig:sig-conf} shows the resulting confusion matrices of the MADELON task when the significance training algorithm is run with 15 training datapoints ($N_t = 15$). For the photonic SNN architecture with 2048 nodes a peak accuracy of 92.5\,$\%$ was obtained and for the 4096 node architecture a peak accuracy of 95.2\,$\%$ was achieved. Again, as expected the larger node number architecture provided the higher classification performance. Comparing this performance to the OLS training method, also run at $N_t = 15$, we find the significance training shows small improvement on peak accuracy in both cases (2048 and 4096 node architectures). This improvement to performance shows that despite the same training set size, by considering only nodes significant to specific classes, we can achieve similar or higher classification performance. Furthermore, it is important to highlight that the training node number used in the significance scoring method was substantially lower than that used in the OLS method. In the novel training method only the top 20 most significant nodes ($N_n=20$) were used in both the reported SNN architecture cases (2048 and 4096). The novel training method therefore only uses 1$\%$ and 0.5$\%$ (for 2048 and 4096 nodes respectively) of the total nodes used for OLS training where every node in the network is trained ($N_n=N_v$). This result indicates that training less nodes overall is not only more efficient computationally, but does not directly hinder the  performance of the system (in this case increase classification performance). These newly presented classification performance values are not however the limit of the photonic SNN with this training scheme. In Figs.\,\ref{fig:perf_test_2} and \ref{fig:perf_test_4}, we consider a larger range of training set sizes ($N_t$), and compare their optimal performance as well as their optimal training node number ($N_n$).  

Figures\,\ref{fig:perf_test_2}\,(a) and \ref{fig:perf_test_4}\,(a) show the peak accuracy achieved in the MADELON classification task as a function of the training set size used, for both 2048 and 4096 network architectures, respectively. Figures\,\ref{fig:perf_test_2}\,(b) and \ref{fig:perf_test_4}\,(b) show in turn the optimal training node number ($N_n$) used to attain the shown optimal accuracy.

Both Figs.\,\ref{fig:perf_test_2} and \ref{fig:perf_test_4} reveal two key results. First, the photonic SNN achieves very high accuracy levels despite the complexity of the MADELON task and its very large number of features. Specifically, a maximum  accuracy of 94.4$\%$ is achieved for the case of a 2048 node architecture, and a maximum accuracy of 95.7$\%$ is achieved for the case of a 4096 node architecture. The performance of this photonic SNN system is in fact higher than, the results of previous software-based machine learning algorithms reported following the NIPS 2003 challenge \cite{GUYON20071438}, which reached accuracy levels of up to 93.78 $\%$. Importantly, one key difference here is that a photonic SNN is applied to the MADELON task with a very hardware friendly implementation (using just one VCSEL), low-power operation (sub-pJ energy per spike, $\sim$\,150\,$\mu$W average optical powers, and 3.5\,mA of applied bias current), and ultrafast performance (250\,ps/node yielding a total processing time of 512\,ns and 1024\,ns per datapoint, for 2048 and 4096 nodes respectively).
Secondly, the performance of the photonic SNN increases with the training set size (associated with best-fit training), and the optimal value of $N_n$ may slightly decrease with training set size. This result is somewhat expected as training with more datapoints typically improves system performance, resulting in a lower dependence on high node numbers. Thirdly, Figs.\,\ref{fig:perf_test_2} and \ref{fig:perf_test_4} show that the optimal training node number ($N_n$) remains consistently low while the system retains very high performance, across all training set sizes ($N_t$) (from 1 to 100). Only small node numbers, as low as $<$\,10 nodes (out of the total node count $N_v$) need to be considered to successfully achieve high performance in the MADELON task. This means that independently of the training set size the high reduced training requirements of the significance scoring method remain a key benefit to the alternative training scheme. Finally, according to Figs.\,\ref{fig:perf_test_2} and \ref{fig:perf_test_4}, very small training set sizes, $N_t\leq\,10$ ($\leq$\,3\,$\%$ of the total dataset inputs), are capable of effectively training the photonic SNN system and achieving high accuracy ($>90\%$) in the MADELON task. Comparing the previous OLS results for a network architecture of 4096 (from \ref{Sect:2.1}) to those of Fig.\,\ref{fig:perf_test_4}, we find that at training set sizes $<$\,15, the significance scoring method achieves similar levels of classification accuracy. Making use of the novel training approach can therefore enable similar performance to the OLS method but with the remarkable benefit of being more flexible in the number of training datapoints.

\section{Conclusion}
\label{Sect:3}

This work demonstrates the high classification performance of a novel laser-based photonic spiking neural network in tackling a highly complex, multivariate, non-linear classification task (MADELON) with 500 datapoint features. Importantly, this work also introduces a novel 'significance' training approach that makes use of binary weights (0 or 1), and that leverages on the advantages of the discrete optical spiking signals found in the photonic SNN. The experimental approach to an SNN combines the spiking dynamics of a VCSEL with a novel network architecture inspired by the reservoir computing paradigm to process data entirely optically at very high speed (GHz rates). The SNN uses all-optical neuron-like spikes to create a time multiplexed feed-forward spiking neural network, in which the values of each time-multiplexed (virtual) node are linked through the VCSEL's non-linear temporal dynamics. The computational power of the photonic SNN is demonstrated first using an OLS method of weight training. We show that by training the output layer weights with this OLS approach we can achieve a high accuracies up to 91$\%$ and 94.4$\%$ for the MADELON classification task using SNN architectures of 2048 and 4096 nodes.

Additionally, in this work, we introduce a new 'significance' training approach, which assigns binary weights to (optical spiking) nodes according to their overall usefulness/significance score. In this approach, only high significance scoring nodes, nodes that spike frequently for one class but not others, are considered and used for the network training and performance evaluation. We show that only a very small fraction ($\leq$\,1$\%$ in the presented case) of the total number of nodes (in the output layer) are required to successfully classify data. We show that classification accuracies of 94.4$\%$ and 95.7$\%$ can be achieved by this new training method. The accuracies provided by the significance training approach show an improvement on those achieved by the OLS method, while also significantly reducing the number of training nodes. Moreover, the photonic SNN demonstrated classification performance that improved upon the benchmark accuracy (93.78$\%$) achieved by software-implemented NNs reported by the dataset authors \cite{GUYON20071438}. Additionally, we demonstrated that the photonic SNN, trained with the new significance method, can also realise high level performance with small training set sizes ($<$10 datapoints), further reducing the overall resources necessary for training the optical system. 

Finally, the presented photonic SNN also offers several inherent physical and computational benefits over traditional digital semiconductor processing systems, notably ultrafast performance (250\,ps/node), low-power usage ($\sim$\,150\,$\mu$W average optical powers, and 3.5\,mA of applied bias current)  importantly, a hardware friendly implementation (using just one VCSEL to process all virtual nodes). Furthermore, the VCSEL-based photonic SNN can adjust performance and processing rate by changing the number of virtual nodes used in the system, which can be done by arbitrarily and on the fly during pre-processing. In conclusion we believe these results open possibilities for further photonics-based processing systems that run and operate entirely on optical hardware, and that are capable of solving highly complex tasks with very high accuracy and ultrafast, energy-efficient operation.

\section*{Acknowledgments}


\subsection*{Author Contributions} 
D. Owen-Newns performed the pre- and post-processing of all data. Both D. Owen-Newns \& J. Robertson performed the experimental runs of the photonic system. A. Hurtado supervised all research efforts. All authors contributed equally to the writing of the manuscript. 





\subsection*{Funding}
The authors acknowledge this work was supported by the UKRI Turing AI Acceleration Fellowships Programme (EP$/$V025198$/$1), by the European Commission (Grant 828841-ChipAI-H2020-FETOPEN-2018-2020), and by the UK EPSRC (EP$/$N509760$/$1, EP$/$P006973$/$1).




\subsection*{Conflicts of Interest}
The author(s) declare(s) that there is no conflict of interest regarding the publication of this article.



\subsection*{Data Availability}
All data underpinning this publication are openly available from the University of Strathclyde KnowledgeBase at https://doi.org/x.xxxxxxx. For the purpose of Open Access, the author has applied a CC BY public copyright licence to any Author Accepted Manuscript (AAM) version arising from this submission.

\printbibliography

\end{document}